\documentclass[fleqn,usenatbib]{mnras}

\usepackage{newtxtext,newtxmath}

\usepackage[T1]{fontenc}

\DeclareRobustCommand{\VAN}[3]{#2}
\let\VANthebibliography\thebibliography
\def\thebibliography{\DeclareRobustCommand{\VAN}[3]{##3}\VANthebibliography}

\usepackage{graphicx}	
\usepackage{amsmath}	
\usepackage{xcolor}  

\newcommand{\msun}{M$_{\odot}~$}

\newcommand{\dgr}{$^{\circ}~$}

\title[The Magellanic Puzzle]{The Magellanic Puzzle: origin of the periphery}

\author[P. Massana et al.]{
Pol Massana,$^{1}$\thanks{E-mail: pol.massana@montana.edu}
David L. Nidever$^{1}$,
and Knut Olsen$^{2}$
\\
$^{1}$Department of Physics, Montana State University, P.O. Box 173840, Bozeman, MT 59717-3840, USA \\
$^{2}$NSF's National Optical-Infrared Astronomy Research Laboratory, 950 N. Cherry Ave., Tucson, AZ 85719, USA 
}

\date{Accepted XXX. Received YYY; in original form ZZZ}

\pubyear{2023}

\begin{document}
\label{firstpage}
\pagerange{\pageref{firstpage}--\pageref{lastpage}}
\maketitle

\begin{abstract}
In this paper, we analyse the metallicity structure of the Magellanic Clouds using parameters derived from the \textit{Gaia} DR3 low-resolution XP spectra, astrometry and photometry. We find that the qualitative behavior of the radial metallicity gradients in the LMC and SMC are quite similar, with both of them having a metallicity plateau at intermediate radii and a second at larger radii.  The LMC has a first metallicity plateau at [M/H]$\approx$$-$0.8 for 3--7\degr, while the SMC has one at [M/H]$\approx$$-$1.1 at 3--5\degr.  The outer LMC periphery has a fairly constant metallicity of [M/H]$\approx$$-$1.0 (10--18\degr), while the outer SMC periphery has a value of [M/H]$\approx$$-$1.3 (6--10\degr).
The sharp drop in metallicity in the LMC at $\sim$8\dgr and the marked difference in age distributions in these two regions suggests that there were two important evolutionary phases in the LMC.
In addition, we find that the Magellanic periphery substructures, likely Magellanic debris, are mostly dominated by LMC material stripped off in old interactions with the SMC. This presents a new picture in contrast with the popular belief that the debris around the Clouds had been mostly stripped off from the SMC due to having a lower mass. We perform a detailed analysis for each known substructure and identify its potential origin based on metallicities and motions with respect to each galaxy.
\end{abstract}




\section{Introduction}

The Large and Small Magellanic Clouds (LMC and SMC, respectively) are dwarf galaxies that constitute the largest satellites of the Milky Way (MW). The LMC sits at the larger end of the dwarf galaxy spectrum, with a mass of about $10^{11}$\msun \citep{Erkal2019b}, while the SMC is expected to be $\sim10$ times smaller when it was in isolation \citep{Besla2007B}, while the current dynamical mass is thought to be $\sim 10^9$\msun \citep{DiTeodoro2019}. 
Their presence in our vicinity provides an excellent opportunity to study the process of galaxy evolution on relatively small scales. The reason why this is so important, is that mergers between dwarf galaxies are expected to be numerous and are one of the main drivers of galaxy evolution, especially at high redshift. The difficulty in resolving individual stellar populations at the large distances of most dwarf galaxies greatly enhances the value of studying the Magellanic Clouds (MCs). Being roughly $50$ and $60$ kpc \citep{Pietrzynski2019,Scowcroft2016} away from the Sun, they sit well inside the MW halo. For decades now, they have been observed to show evident signs of interaction between them. One of the more striking ones being the Magellanic Bridge \citep{Hindman1961,Hindman1963}, which was discovered as a tail of young stars coming off the SMC and has been long associated to the LMC stripping off its gas due to an interaction. This interaction was later hypothesised in simulations to be a very close and recent passage between the two galaxies \citep{Besla2012}. This was later confirmed using models that incorporated more precise measurements of their velocities, with an impact parameter of $\sim 7.5$ kpc and 150 Myr ago \citep{Zivick2018, Choi2022}. 
Perhaps the most striking feature though is the Magellanic Stream (MS; \citealt{Putman1998,Nidever2008}), a trail of HI gas extending over 200\dgr across the sky \citep{Nidever2010}. It has both a trailing and a leading component, called the Leading Arm (LA).  After a decades-long search for a predicted stellar counterpart to the Magellanic Stream  \citep[e.g.,][]{Philip1976,Brueck1983,Kunkel1997,Guhathakurta1998}, there is
recent evidence that one might have been found \citep{Zaritsky2020}.
In addition, a recently-formed stellar association (117 Myr) was found that was born in the LA gas
\citep{Price-Whelan2019,Nidever2019}.
Part of the challenge of detecting the stellar counterpart may be because the distance to the trailing part of the Stream is quite uncertain,
with some simulations giving wildly different distance suggestions for it \citep{Lucchini2020, Lucchini2021}. 

Other results from simulations have long predicted the existence of tidal debris extending far outside their main bodies \citep[e.g.,][]{Besla2007B,Diaz2012,Besla2012}. For this reason, the outskirts of the MCs have been a subject of intense scrutiny over the last decade. 
Deep photometric surveys, such as the Dark Energy Survey \citep{DES2016}, \textit{Gaia} \citep{Gaia2016}, SMASH \citep{Nidever2017,Nidever2020a}, or VMC \citep{Cioni2011}, have enabled the discovery of many different stellar structures around the MCs \citep{Belokurov2019,Gaia2021MC,ElYoussoufi2021}. These features are able to yield crucial information on the timing of the interactions. So far, most of the studies for these features have focused on modelling their morphology and kinematic information on a limited number of stars \citep{Cullinane2020,Cheng2022}. To determine the ages of each structure and their origins, a full understanding of their metallicities would provide a much needed constraint.

Another avenue to explore the history of a galaxy are metallicity gradients. It is expected that, among other factors, the influence of satellites around a galaxy can cause stars to migrate radially \citep{LyndenBell1972,Sellwood2002}, changing the metallicity distribution function (MDF) at different radii. This has been observed for large galaxies like the MW \citep{Hayden2015,Loebman2016}, however studies in other galaxies, specially dwarfs, are a challenge due to their distance and size. Even for the MCs, large scale metallicity studies, based on spectroscopic data, have historically had a limited scope. The large apparent size of the two galaxies on the sky have so far made the task very time consuming, requiring many hours of observation \citep{Nidever2021a}. 

In this paper, we use the largest metallicity catalogue released to date for the Magellanic Clouds, enabling a completely new approach to galactic archaeology outside of the MW disk. The \textit{Gaia} mission has proven to be extremely successful in the study of the MW as well as the MCs. Thanks to its all-sky coverage and astrometric information, it is capable of separating MW and Magellanic member stars with unprecedented precision. Its third data release included a batch of 220 million low-resolution spectra \citep{GaiaDR32022} in its BP and RP bands, down to $G$=17.65. Unfortunately, the physical parameters that were released with it suffered from several systematic uncertainties that prevented its full exploitation, especially for the MCs. This prompted the community to come up with novel ways to exploit the raw spectra in order to obtain alternative values for important physical parameters such as metallicity, surface temperature and gravity. In this paper, we choose to utilise the recent public release of a set of physical parameters based on \textit{Gaia} spectra, astrometry as well as additional infrared photometry from other surveys. The parameters have been obtained using machine learning algorithms and the methodology is explained in \cite{Andrae2023}. Despite the compromise taken by being limited to $G\leq$17.65,
this is still a very compelling catalog due to the full MC coverage and ability to make a very accurate MC selection, helped by adding surface gravity information.

This paper is structured as follows. In Section \ref{section:data}, we discuss the data products used in our analysis.  This follows with a presentation of our results in Section \ref{section:results}.  The implications of our findings for the broader Magellanic context are discussion in Section \ref{section:discussion}.  Finally, Section \ref{section:conclusions} lays out our conclusions.

\section{Data}
\label{section:data}

\subsection{Metallicity data} \label{subsec:a23_data}

In this study, we use a combination of \textit{Gaia} astrometric and photometric data \citep{GaiaDR32022} with recently published metallicity information, also based on \textit{Gaia} XP spectra \citep[][hereafter A23]{Andrae2023}. The spectroscopic information includes metallicity, effective temperature and surface gravity values. The full dataset from A23 includes 175 million objects across the full sky, going down to $G=17.65$, with a few additional sources at deeper magnitudes. These are improved values from the official \textit{Gaia} DR3 values \citep{DeAngeli2022} that are only based on XP spectra and had several known shortcomings. The improvement mainly comes from adding infrared photometry from AllWISE \citep{Cutri2021} and parallax information into a decision tree machine learning algorithm. Unfortunately, it also comes at the loss of about 45 million sources from the original XP spectra release in \textit{Gaia} DR3 due to the low completeness of the AllWISE data. A23 use XGBoost \citep{Chen2016a} trained on APOGEE DR17 \citep{Abdurrouf2022} data and very metal poor stars from \cite{Li2022}. We refer the reader to the original paper for more details on the methodology.

\begin{figure}
    \centering
    \includegraphics[width=\columnwidth]{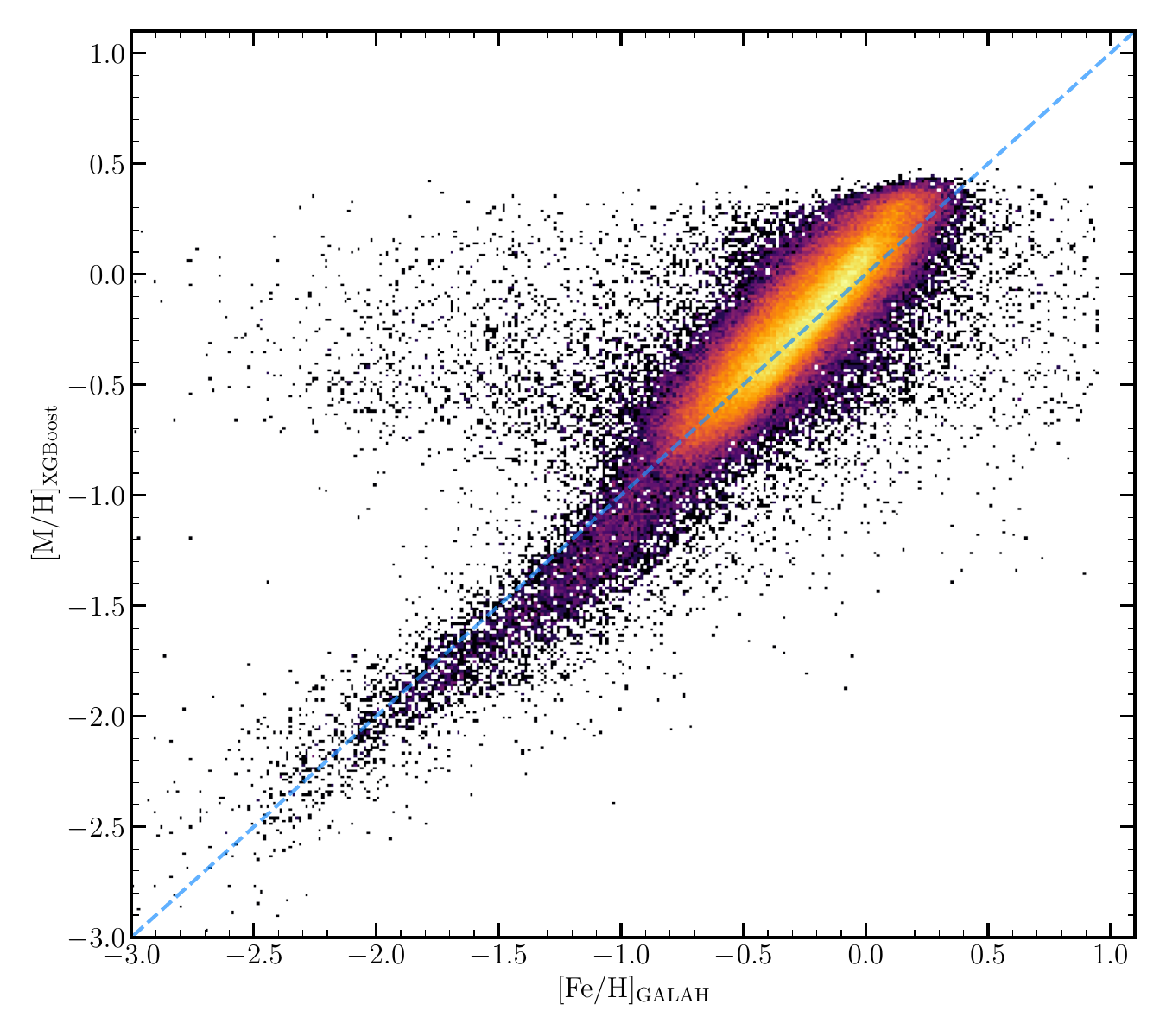}
    \caption{Comparison between metallicity values from GALAH DR3 ([Fe/H]) and those derived by XGBoost using \textit{Gaia} DR3 XP spectra ([M/H]) \protect\cite{Andrae2023}. The comparison is limited to giant stars with $\log(g)\leq3.5$ and T$_{\mathrm{eff}}\leq5200$K using XGBoost values.}
    \label{fig:galah_vs_gaia}
\end{figure}

In order to assess the reliability of the A23 values, we perform a cross-match with GALAH DR3 \citep{Buder2021} data, which is a southern hemisphere survey that also partly covers the MCs. We note that a similar comparison is already made in the A23 release paper, with the full GALAH sample. In our case, however, we limit the comparison to stars outside the plane of the MW ($|b|>10^{\circ}$) 
and include only giant stars with A23 values $\mathrm{log}g>3.5$, T$_{\mathrm{eff}}\leq5200$K and BP-RP$>0$. This is done because the GALAH data is only deep enough to capture giant stars in the Clouds, and these are the type of stars we will use for our analysis. As shown in Figure \ref{fig:galah_vs_gaia}, the A23 [M/H] values usually stay within 0.25 dex of the GALAH values, except within the -1 to -1.5 range where the GALAH [Fe/H] are systematically higher. Any differences between the A23 metallicities and those from GALAH might well be due to different calibrations between GALAH and APOGEE (the main training sample for XGBoost). It is therefore worth keeping in mind this small discrepancy when comparing our results to GALAH in the future.

\subsection{Magellanic Cloud Membership Selection} \label{subsec:mcs_selection}

Because the A23 data does not come with the full set of \textit{Gaia} columns, in order to select Magellanic members within the A23 data, we combine it with parallax, proper motion and colour-magnitude diagram (CMD) cuts using the full \textit{Gaia} DR3 set. To obtain the \textit{Gaia} data, we limit the search to outside the MW plane again ($|b|>10^{\circ}$) 
, stars with \texttt{has\_xp\_continuous=True}, $G$$\leq$17.65 and ${BP}-{RP}$$>$0. The $G$-band limit is to avoid the few fainter sources, that were released as part of DR3, but are not interesting to our science (e.g., quasars, white dwarfs). The colour limit is because we also make a separate CMD cut later on that already includes this limit, and it makes the search more manageable. 

\begin{figure}
    \centering
    \includegraphics[width=\columnwidth]{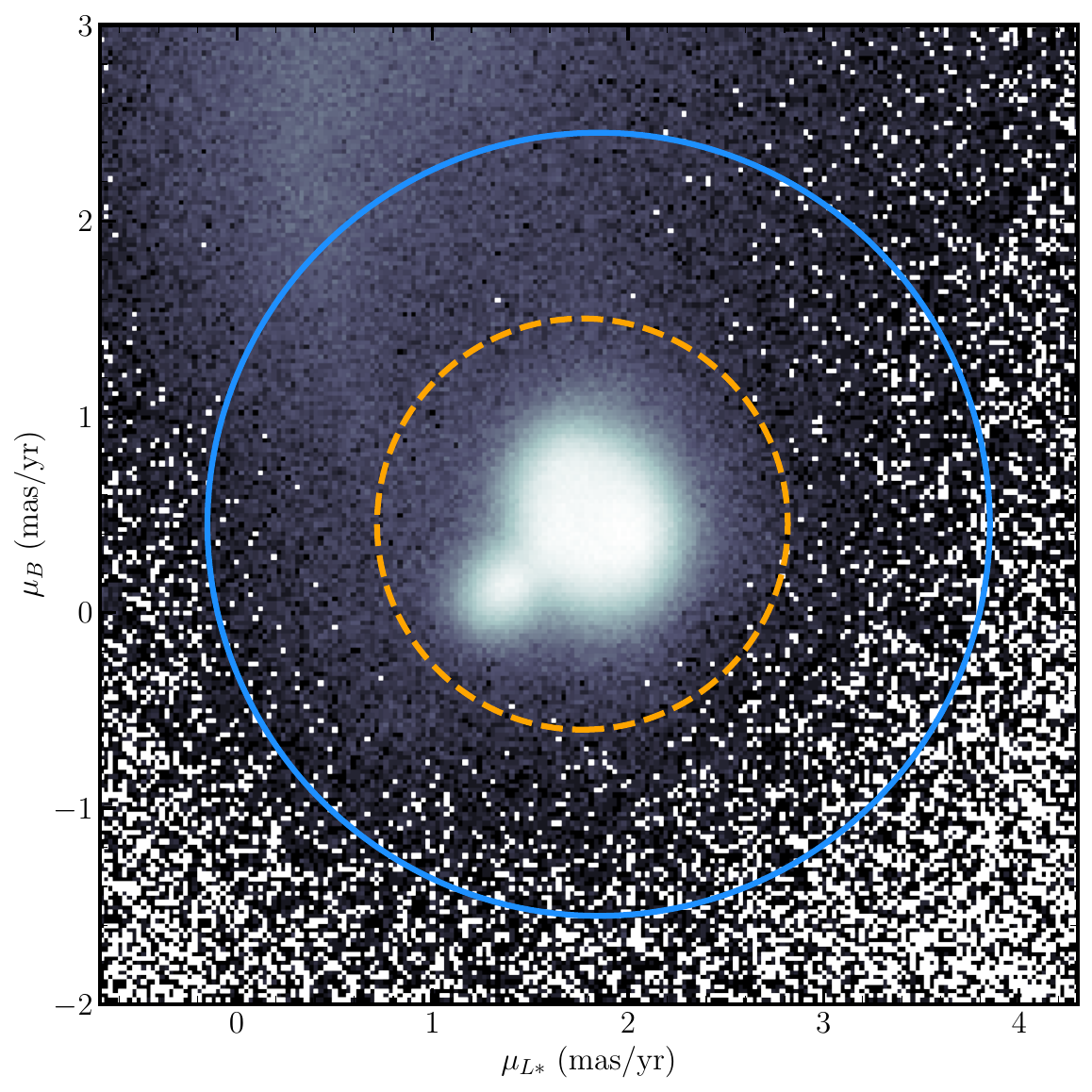}
    \caption{Proper motion distribution, in Magellanic Stream coordinates, for stars cross-matched between the A23 metallicity data and the full \textit{Gaia} DR3 catalogue. Only stars that are within 40 degrees of the LMC and 30 degrees from the SMC are shown. The blue solid circle indicates the cut for the metallicity sample, that we use for most of our scientific analysis in this paper. The dashed orange line is used to enhance the view of periphery structures in Figure \ref{fig:periph_mdfs}, and it is applied to the whole \textit{Gaia} DR3 set (see Section \ref{subsec:periphery_results} for details).}
    \label{fig:pm_cut}
\end{figure}

Once the cross-match between the A23 data and the full \textit{Gaia} DR3 data is made, we perform astrometry and photometry cuts on our metallicity sample. We decided for a conservative approach, because of its bright magnitude limit and because we can refine our membership selection using spectroscopic information later. We keep only stars with a parallax consistent with being at distances larger than 30 kpc with a requirement of $|\varpi|/\sigma_{\varpi} > 10$.
For the proper motions, we pick the central proper motion for the system value to be $(\mu_{\mathrm{L}},\mu_{\mathrm{B}})=(1.85,0.45)$ mas yr$^{-1}$, based on the median for the whole sample, in Magellanic Stream coordinates \citep{Nidever2008},
and use all the stars within 2 mas yr$^{-1}$ of that value (as can be seen in Figure \ref{fig:pm_cut}). To perform the RGB cut, we first use \citet[][hereafter SFD98]{Schlegel1998} to correct for reddening.  We use the following coefficients to transform the $E(B-V)$ values to \textit{Gaia} passbands:

\begin{align*}
    &A_G = 0.731786 \cdot R_V \, E(B-V) \\
    &A_{BP} = 0.93751705 \cdot R_V \, E(B-V) \\
    &A_{RP} = 0.50695229 \cdot R_V \, E(B-V),
    \label{eq:extinction_coeffs}
\end{align*}

\noindent
with $R_V=3.1$ and $E(B-V)$ the SFD98 values. The coefficients are based on the Galactic extinction curve by \cite{Fitzpatrick1999} combined with the \textit{Gaia} passband information\footnote{The passband information can be downloaded from \url{https://www.cosmos.esa.int/web/gaia/edr3-passbands}.} \citep{Riello2021}. The SFD98 values also include the correction proposed in \cite{Schlafly2011} included in the coefficient.
The SFD98 values are known to be unreliable in the central parts of the MCs, as acknowledged by the authors themselves. The $E(B-V)$ values are generally overestimated in these areas, which would result in an incorrect selection of RGB stars. To solve this issue, we manually adjust the values in the central parts. For the area inside 4.5$^{\circ}$ of the centre of the LMC (L$_{\mathrm{MS}}$,B$_{\mathrm{MS}}$)=($-0.6^{\circ}$,$3.6^{\circ}$) and 2.5$^{\circ}$ from the centre of the SMC (L$_{\mathrm{MS}}$,B$_{\mathrm{MS}}$)=($-15.53^{\circ}$,$-11.58^{\circ}$), we multiplied the SFD98 $E(B-V)$ values by 0.07.
Because our analysis and the discussion of this paper is centred around the outer parts of the galaxy and outside this inner areas, this approximation is more than sufficient. We note though, that to study the main body of the galaxies as a whole, this is something that should be addressed more accurately and there are other works in the literature that do this specifically \citep{Choi2018a,Bell2020}.

\begin{figure}
    \centering
    \includegraphics[width=\columnwidth]{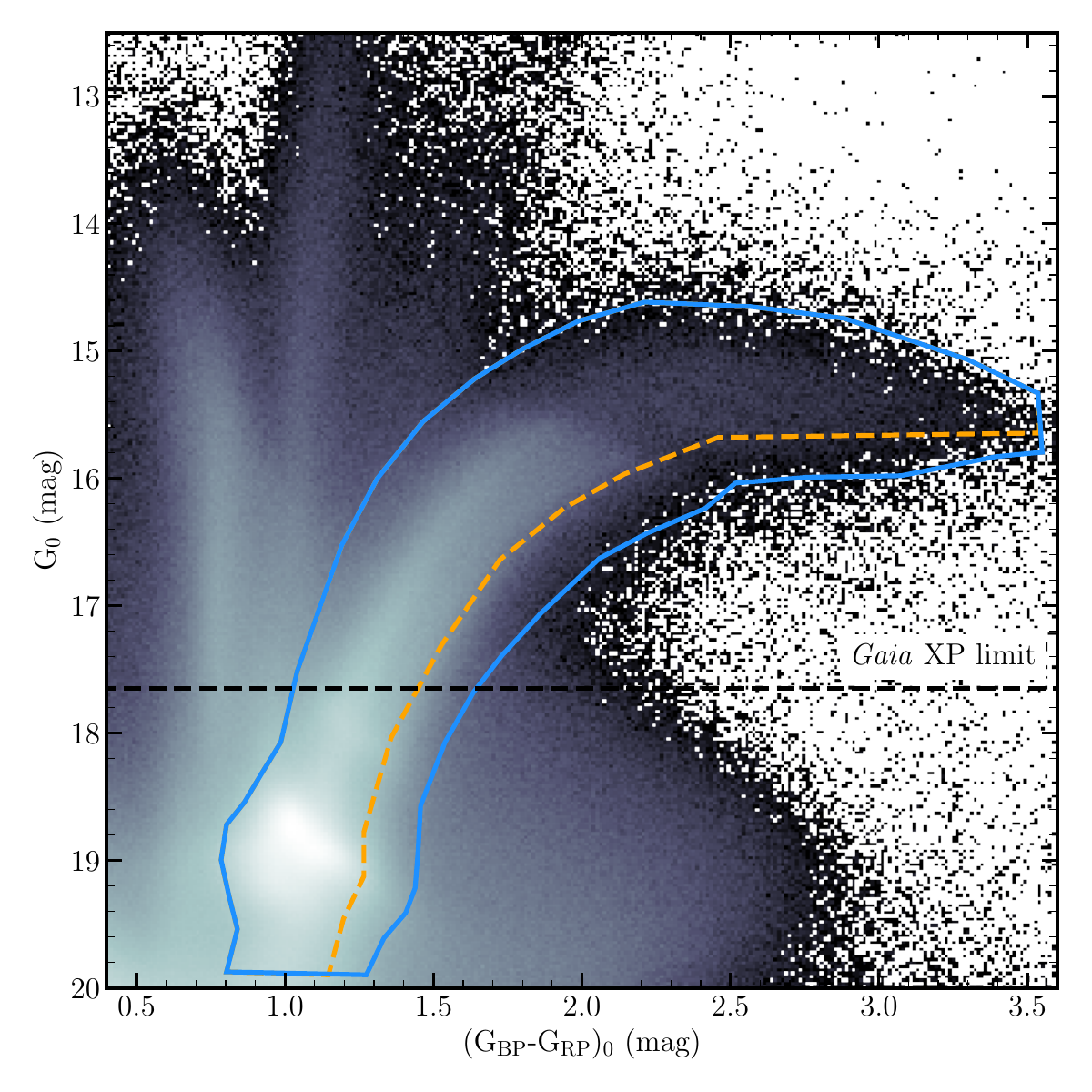}
    \caption{Colour-magnitude diagram of \textit{Gaia} DR3 stars around the MCs (see Section \ref{subsec:mcs_selection} for a description of how it was obtained). The blue polygon shows the red giant branch cut used for the science presented in this paper. The orange dashed line defines a stricter cut made to enhance the contrast of the periphery structures in Figure \ref{fig:periph_mdfs}; it includes only those stars within the blue polygon that fall blue-ward of the orange line. The black dashed line shows the depth limit on \textit{Gaia} XP data, which translates to the cut that our main metallicity sample has.}
    \label{fig:rgb-cut}
\end{figure}

Finally, we perform a CMD cut to select only giant stars down to $G$=17.65 for the main metallicity sample, shown in Figure \ref{fig:rgb-cut}. This is later improved using the added information of the $\log g$ values coming from the spectroscopy data, by using a $\log g \leq 3.5$. This helps clean the MW foreground because most of the closer stars are dwarfs or main-sequence stars. The final sample with metallicity data contains 471,682 rows and it is all-sky, although in this work we will focus on the Magellanic Clouds area.

\subsection{Kinematic model for center of mass motion} \label{subsec:knut_models}

In order to compare the motion of the periphery substructures to the internal motion of each of the Clouds and help distinguish their origin, we need to account for the fact that the bulk motion vectors of the Clouds project into the tangent plane and line of sight in ways that depend on location with respect to the centre of mass (COM), as shown by \citet{vdM2002}. Our main concern is with the metal-poor structures and their relation with either of the galaxies. Although in our analysis we use both COM motions (LMC and SMC), one is enough to differentiate their potential origin. We find that the SMC model is able to separate each structure more clearly, and so we will use it to drive most of our discussion later.
For the SMC, we use the COM motion from the best-fit model of \citet{Zivick2021}, and use the formalism of \citet{vdM2002}, as implemented by \citet{Choi2022}, to compute the COM motion contribution at the positions of our stars.  We use these computed contributions to place our measured kinematics in the SMC COM reference frame. 

\begin{figure*}
    \centering
    \includegraphics[width=0.8\textwidth]{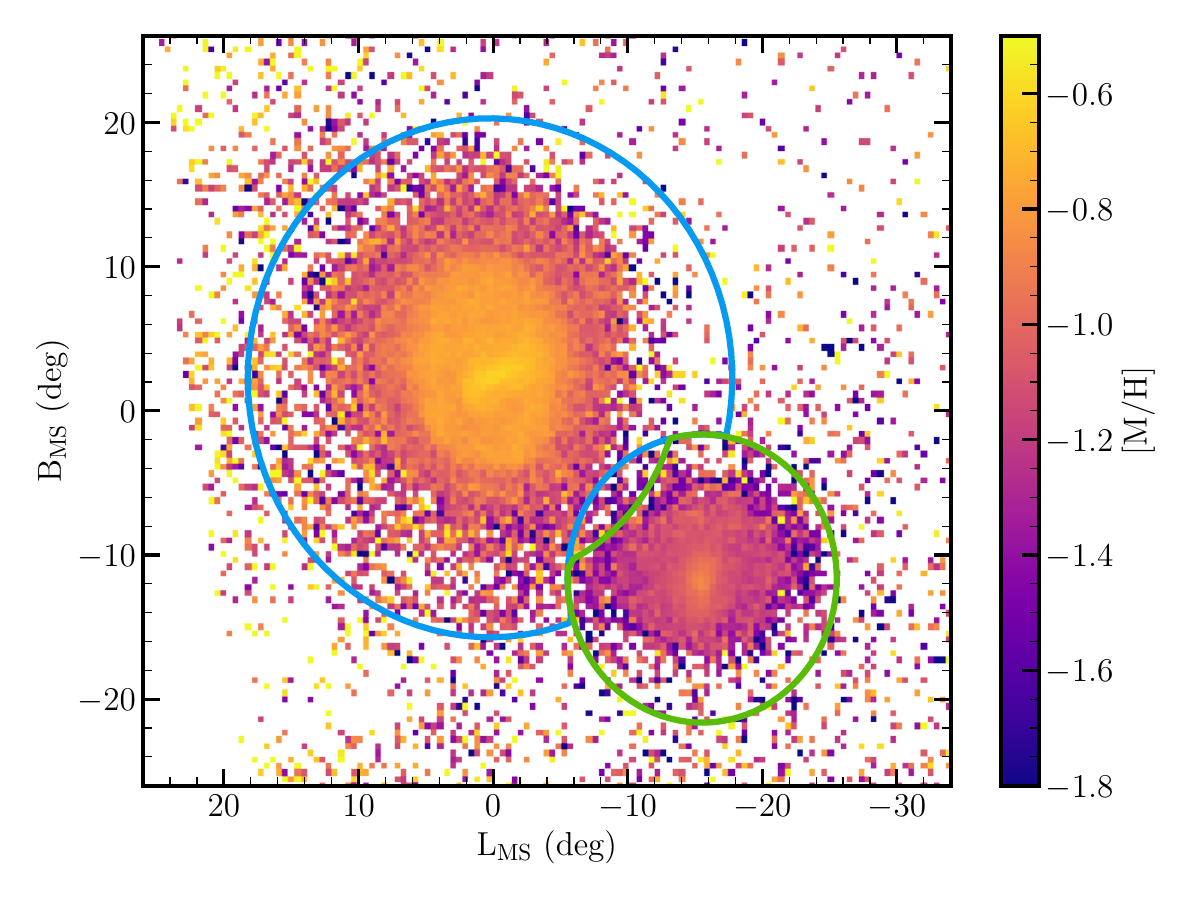}
    \caption{Gaia XPSpec metallicity map of the Magellanic Clouds. Proper motion, parallax, CMD and surface gravity cuts have been applied to the abundance data. See Section \ref{subsec:mcs_selection} for more details on the selection. The LMC shows significant more metal enrichment than the SMC, which is in line with the known patterns observed in the literature. The blue and green outlines around the LMC and SMC are our selection areas to calculate the metallicity gradients showcased in Figure \ref{fig:radgradients}.}
    \label{fig:mhmap}
\end{figure*}

\section{Results}
\label{section:results}

\subsection{Metallicity Map}
\label{subsec:metalmap}

Figure \ref{fig:mhmap} shows the metallicity map across the Magellanic Clouds.  It is clear that the LMC has a higher metallicity than the SMC, on average.  Also, radial metallicity gradients are clearly present in both the LMC and the SMC.  The metallicity in the outer LMC appears to plateau at [M/H] $\approx$ $-$1.0, while the outer SMC reaches values of [M/H] $\approx$ $-$1.5.  While there is much scatter in the metallicities of the far outer Magellanic periphery, some patterns are apparent there as well.  The mean metallicities of the northern, eastern and southeastern periphery of the LMC look similar to the mean metallicity of the outer LMC at [M/H] $\approx$ $-$1.  In contrast, the periphery regions directly around the SMC look more like the edge of the SMC at a value of roughly [M/H] $\approx$ $-$1.5.  Below, we discuss the metallicity distributions of various substructures in more detail.

\subsection{Radial Metallicity Gradients}
\label{subsec:radgradients}

\begin{figure*}
    \centering
    \includegraphics[width=0.48\textwidth]{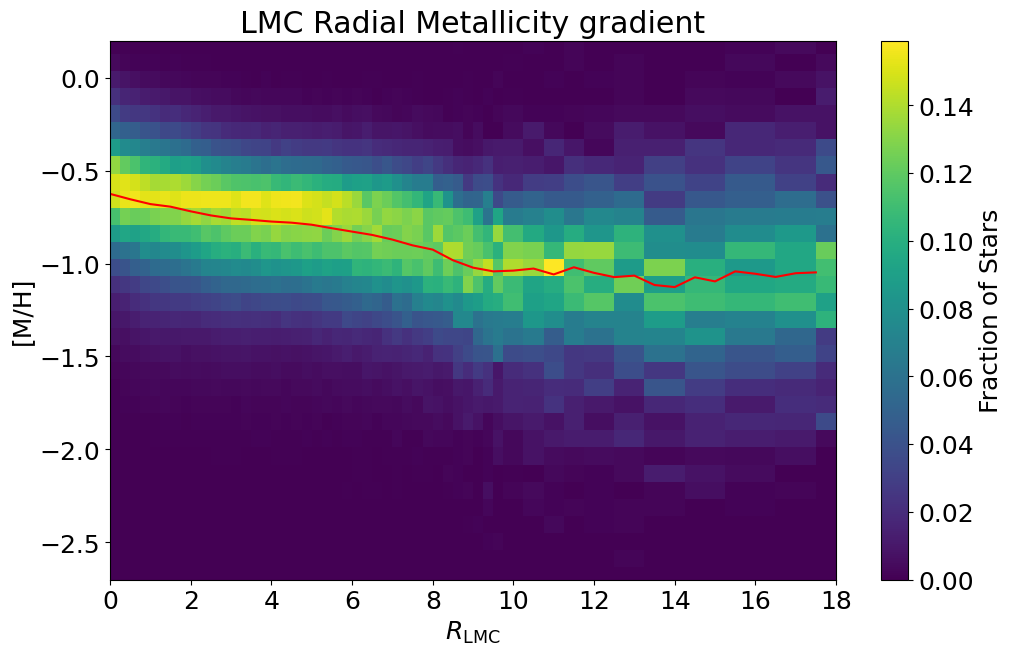}
    \includegraphics[width=0.48\textwidth]{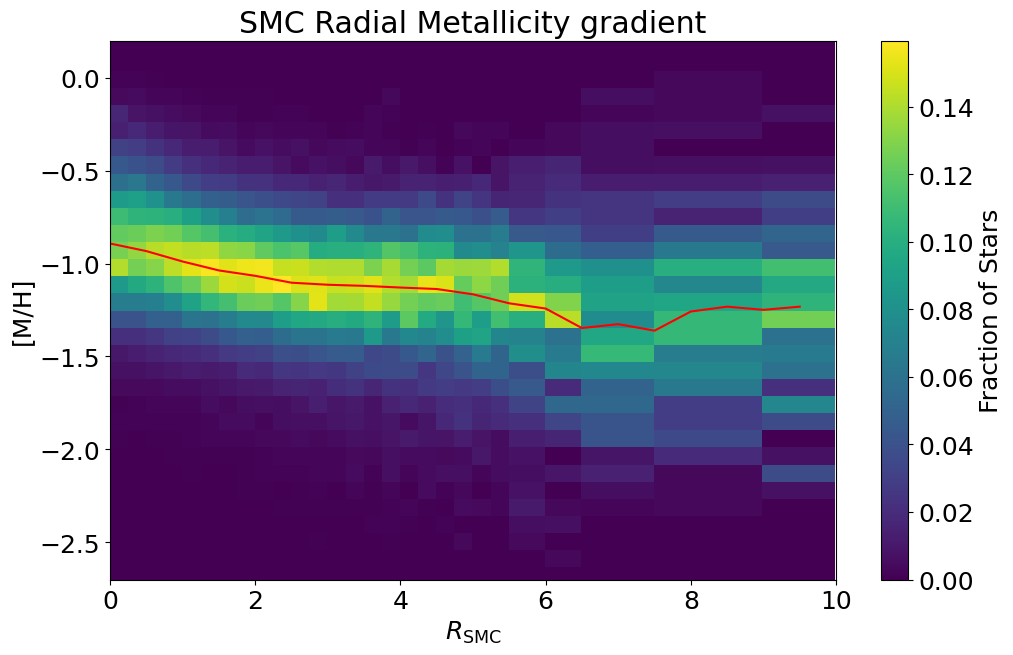}
    \caption{Radial metallicity gradients of the LMC (left) and SMC (right). The background image shows the 2-D density map of stars normalized by the number of stars at a given radius to produce the fraction of stars at each metallicity.  This is effectively a metallicity distribution function (MDF) at each radius.  At larger radii the number of stars decreases and we combine radial bins until there are at least 500/300 stars (LMC/SMC). The red solid line shows the median value in bins of 0.75\degr.}
    \label{fig:radgradients}
\end{figure*}

Figure \ref{fig:radgradients} shows the radial metallicity gradients for the LMC (left) and SMC (right).  The background image shows the normalized 2-D density map which is effectively the metallicity distribution function (MDF) at each radius.  At larger radii some radial bins are combined until there are at least 200 stars.  In addition, the red line shows the median in bins of 0.75\degr. Regions in the Bridge are not included to reduce any cross-contamination.

Both galaxies show quite similar patterns in their radial metallicity gradients.  There is a ``plateau'' in the metallicities at intermediate radius (LMC: [M/H]=$-$0.75 for 2--8\degr; SMC: [M/H]=$-$1.1 for 2--5\degr). At smaller radii, the metallicites climb linearly until they reach a maximum of $\approx$$-$0.6 in the LMC and $\approx$$-$0.9 in the SMC. At larger radii, the metallicity drops quickly at the end of the intermediate plateau region reaching a new plateau in the LMC at [M/H]=$-$1.1 for 9--18\degr.  For the SMC, the metallicity drops to [M/H]=$-$1.35 for 6.5--7.5\dgr but then increases again for the last 2.5\dgr to [M/H]=$-$1.15.

\subsection{Periphery structures}
\label{subsec:periphery_results}

\begin{figure*}
    \centering
    \includegraphics[width=0.9\textwidth]{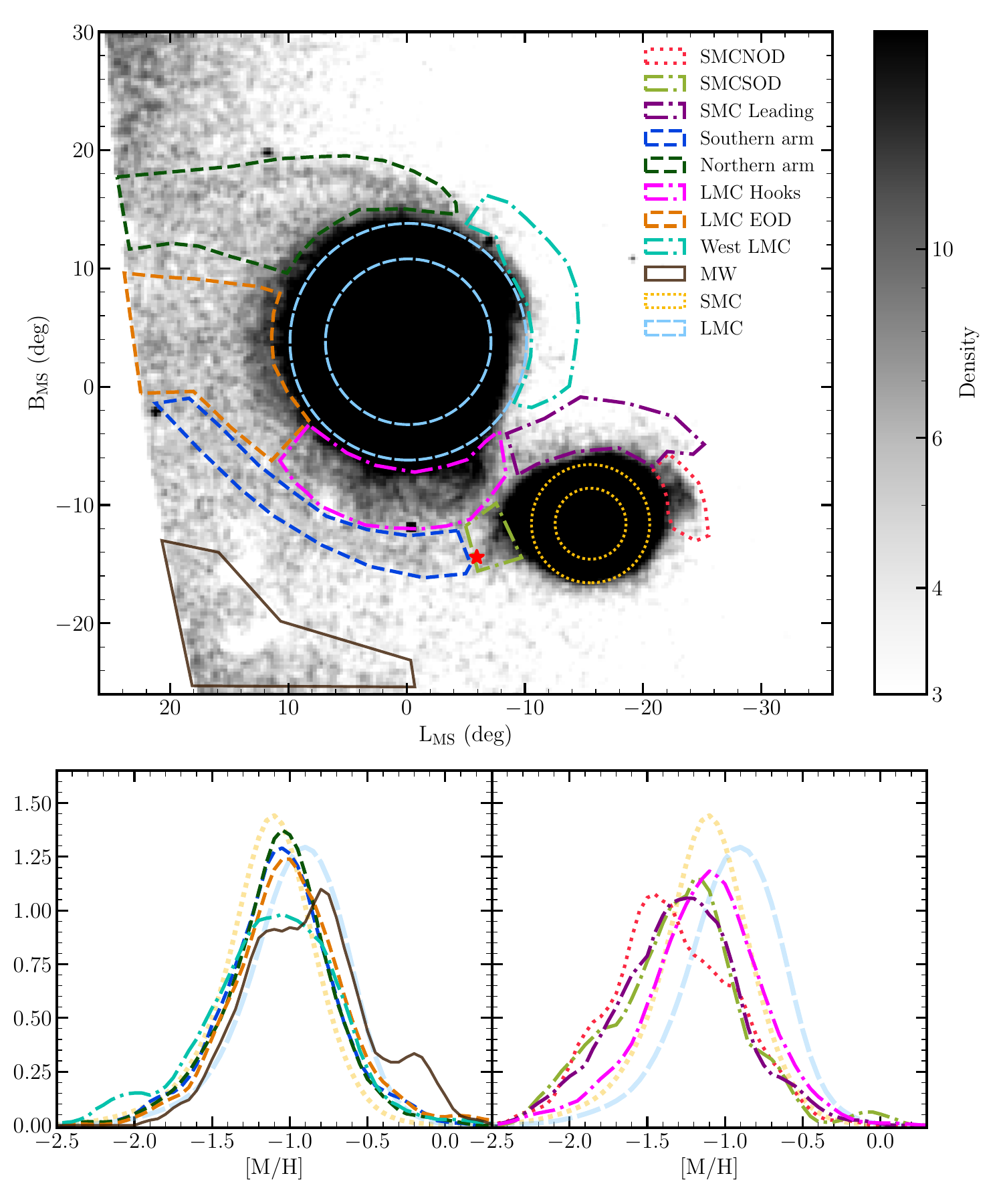}
    \caption{Metallicity distribution functions (MDFs) for Magellanic periphery structures. The top panel shows a density map made using the full data from \textit{Gaia} DR3, it includes cuts in astrometric measurements and a cut in $|b| >10$\degr. The MDFs correspond to a smaller data set which has been cross-matched to the metallicity data and only goes down to $G$=17.65 (see Section \ref{section:data} for more details). Areas that we associate with the LMC are dashed (bottom left), while those we identify as only SMC are dotted (bottom right). Any region where we see evidence for both, or it is unclear, are dash-dotted. The red star inside the SMCSOD area corresponds to Field 3 of the MagES survey \protect\citep{Cullinane2023}, that we discuss in Section \ref{subsec:periphery_results}. Both bottom panels include the SMC and LMC MDFs corresponding to the two ring sections of both galaxies defined in the top panel, to help with the comparison.}
    \label{fig:periph_mdfs}
\end{figure*}

In Figure \ref{fig:periph_mdfs} we show the MDFs of several areas around the outskirts of the Magellanic Clouds. In the top panel, we can see the selected regions in a density map that uses the full \textit{Gaia} DR3 data down to $G$=20 and also with proper motion and parallax cuts applied. These are a little more restrictive for the proper motion side (corresponding to the orange dashed outline in Figure \ref{fig:pm_cut}) than the ones used in the metallicity set, considering that we start from a list of more objects and that we cannot use surface gravity as an additional parameter for decontaminating. The cuts are meant to emphasise the outer structures and get rid of as much contamination as possible. In order to compare the MDFs of the outskirts features and to have a sense of where these could originate from, we select an annular region around the main bodies of the two galaxies. The MDFs of the regions directly around the LMC are depicted in the bottom left panel, while the areas around the SMC (as well as the Southern Arm), are on the bottom right. In Table \ref{tab:periph_mets}, we summarise the main properties of each substructure. Based on previous literature results, we identify the following features:

\begin{itemize}
    \item SMC Northern Over-Density (SMCNOD; red): discovered with Dark Energy Survey \citep[DES:][]{Abbott2018} data as an elliptical overdensity at the edge of the DES footprint \citep{Pieres2017}. It was reported as one of the furthest SMC components ever seen, with a double population of both intermediate and old stars. In our results, this is one of the regions that shows a clearly more metal poor population with a mean [M/H]$=-1.35$ dex. The SMCNOD shows a bi-modality in its MDF which corroborates what \cite{Pieres2017} noted in their discovery of the feature (two distinct populations). An intermediate age population, which we note has the same metallicity as the outer areas of the SMC main body, and an old stellar population that seems to have a significantly lower metallicity at [M/H]$\approx -1.5$ dex.
    \item SMC Southern Over-Density (SMCSOD; light green): this overdensity (first named here as SMCSOD) sits at the opposite direction of the SMCNOD and together with it they are believed to indicate the direction of the tidal disruption in the SMC, which also lines up with its proper motion with respect to the LMC. In the map presented by \cite{Belokurov2019}, this part appears to connect to the Southern arm that we describe below, and, therefore, at the time it reinforced the idea that it was part of an extended tidal debris for the SMC. Recently, \cite[][C23 hereafter]{Cullinane2023} used spectroscopic data to explore a field in this area and determined that it had a double population with two completely different kinematic signatures. They dubbed them ``bulk'' and ``offset'', and associate them both to the SMC, although they note the ``offset'' population is heavily perturbed. Our metallicity results also show a double peak similar to the SMCNOD. Although we will attribute most of the debris here to the SMC based on our metallicity results, the kinematic information reveal a small contribution of LMC debris that can also be seen as a small bump in the MDF at the higher metallicities ([M/H]$\sim -0.7$ dex). 
    \item SMC Leading (purple): this sparsely populated area has been previously observed to contain Magellanic debris, but few times has been named and it has yet to be studied in detail. We note that part of this debris was named by \cite{ElYoussoufi2021} as ``Northern Substructure 2'', but was not analysed further. We name this area ``Leading'' because it is known to be the direction the SMC is moving towards. It also has a very metal poor MDF similar to the other SMC areas, but our selection box could include some LMC material based on the proximity to the galaxy.
    \item Southern Arm (dark blue): although it may look like an extension to the SMCSOD, it has not been associated with it. \cite{Belokurov2019} noted that the debris in this area did not show a strong SMC connection. However, the stars in this area pass through the South Celestial Pole and some extra care needs to be taken to interpret the kinematical signatures. Recently, two independent studies using APOGEE data \citep{Cheng2022,Munoz2023} investigated two fields in the region we highlight here. Their studies show that the prevalence of metal poor stars within the Arm increases with proximity to the SMC, indicating that the Arm may include some SMC material. In our selection region we find mostly LMC material based on the MDF, but some metal poor stars also show up, making it a compatible result with the APOGEE data. In Section \ref{section:discussion} we discuss a connection with the SMCSOD region that we defined and how some material from the SMC might be appearing in this region and, vice-versa, how LMC material is showing up inside SMCSOD.
    \item Northern Arm (dark green): This is an LMC structure identified by \cite{Mackey2016} using deep photometry from Dark Energy Camera \citep[DECam;][]{Flaugher2005}. It has been extensively studied, recently by the MagES Survey \citep{Cullinane2022a}, to confirm if it is a perturbation of the LMC disc. \cite{Cheng2022} and \cite{Munoz2023} both studied this area using APOGEE data, which gave both metallicity and radial motion data. Their results for this area are compatible with each other, but we note that their N1 area (located close to the Carina dSph) has a more metal poor component at [Fe/H]$\sim -1.3$ that is not visible in our MDF. We speculate that this might be due to the smaller number of those metal poor stars compared to our full region, which is much larger than their APOGEE fields. Regardless, our metallicity results put this feature as one of the more metal rich ones, making it a clear connection with the LMC, which is the same connection that all the literature references make.
    \item LMC Hooks (magenta): this area of the Magellanic periphery was first shown to contain tidal debris by \cite{Mackey2018} using DECam data. Later, the view was expanded using \textit{Gaia} DR2 data by \cite{Belokurov2019}. It has been traditionally associated with the LMC. This is another area for which there are published results with APOGEE data in a couple of small regions by \cite{Cheng2022} and \cite{Munoz2023}. Their data support a feature mostly comprised of LMC material based on their MDF. In the data presented in this paper, this region is more metal poor than the main LMC debris areas, with a mean around 0.1 dex lower. We hypothesise that this is due to a mix of LMC and SMC debris present, which is also consistent with the APOGEE results.
    \item LMC Eastern Over-Density (EOD; orange): this diffuse debris was first clearly observed using \textit{Gaia} EDR3, in one of the papers showcasing the science cases \cite{Gaia2021MC}, and it is also attributed as LMC debris. Along with the Northern arm, this is also a higher metallicity area that we associate to the LMC.
    \item West LMC (turquoise): even though this region has not been previously noted to have any prominent structure of debris, we add for completion and because there are enough stars that it would make statistical sense to calculate the metallicity mean and dispersion. It is dominated by LMC debris and a metal-poor component, possibly with SMC origin.
\end{itemize}

\begin{figure*}
    \centering
    \includegraphics[width=0.8\textwidth]{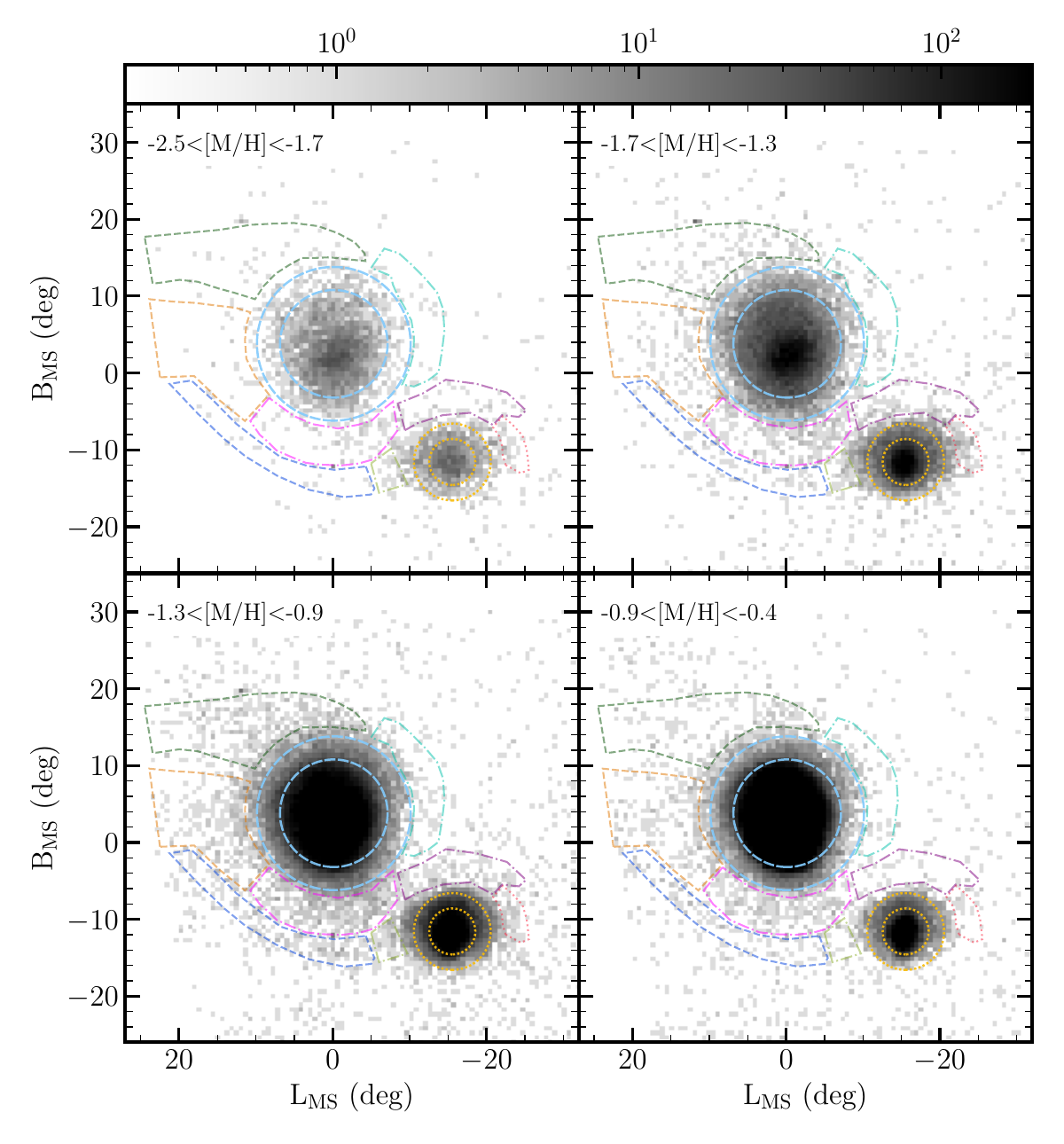}
    \caption{Density map of Magellanic Cloud stars in several metallicity ranges that emphasise different periphery structures. The color scheme and naming of the structures is the same one we used in Figure \ref{fig:periph_mdfs}. Most of the LMC structures are more prominent in the range $-1.3<[M/H]<-0.9$, and the SMC structures appear more clearly in the $-1.7<[M/H]<-1.3$ range.}
    \label{fig:4density_maps}
\end{figure*}

In order to highlight the presence of the substructures in different metallicity bins, we showcase in Figure \ref{fig:4density_maps} four density maps of the metallicity sample. As expected, most of the debris is seen around the LMC in the range $-1.3<\mathrm{[M/H]}<-0.9$, e.g., with a well defined Northern arm. Additionally, the SMC debris is already clearly observed in the more metal poor $-1.7<\mathrm{[M/H]}<-1.3$ range. We point out that the SMC looks very elongated in the most metal poor regime, likely feeling the long-term effects of the orbit around the LMC. The LMC Hooks area has a metal poor component in the MDF presented in Figure \ref{fig:periph_mdfs}.  However, its spatial distribution shows no clear signs of a coherent structure and the debris is far from the SMC. A larger depth and more stars are needed to make out any substructure. On the other hand, there is a clearer structure that connects to the SMC on the most metal rich range. 

\begin{table*}
    \centering
    \caption{Metallicity properties of the main periphery substructures of the MCs highlighted in Figure \ref{fig:periph_mdfs}, calculated using low-resolution \textit{Gaia} spectroscopic results published in \protect\cite{Andrae2023}.}
    \begin{tabular}{l|c|c|c|c}
         Structure & $\mu_{\mathrm{[M/H]}}$ (dex) & FWHM$_{\mathrm{[M/H]}}$ (dex) & N$_{\mathrm{stars}}$ & Origin \\
         \hline
         SMCNOD & -1.34 & 0.90 & 157 & SMC \\
         SMCSOD & -1.30 & 0.69 & 150 & SMC+LMC \\
         SMC Leading & -1.30 & 0.84 & 286 & SMC+LMC \\
         Southern arm & -1.06 & 0.68 & 310 & LMC \\
         Northern arm & -1.07 & 0.61 & 625 & LMC \\
         LMC Hooks & -1.14 & 0.75 & 909 & LMC+SMC \\
         LMC EOD & -1.01 & 0.71 & 596 & LMC \\
         West LMC & -1.12 & 0.93 & 221 & LMC+SMC \\
         MW & -0.88 & 0.80 & 114 & - \\
         SMC & -1.14 & 0.30 & 10,182 & SMC \\
         LMC & -0.96 & 0.31 & 19,712 & LMC \\
         \hline
    \end{tabular}
    \label{tab:periph_mets}
\end{table*}

\subsection{Kinematic properties of structures} \label{subsec:pms_periph}

In order to identify potential origins of the substructures, we complement our metallicity results with a kinematical analysis. We use the model of the MCs described in Section \ref{subsec:knut_models} to calculate the proper motions of each star with respect to the LMC and SMC COMs. 

\begin{figure}
    \centering
    \includegraphics[width=\columnwidth]{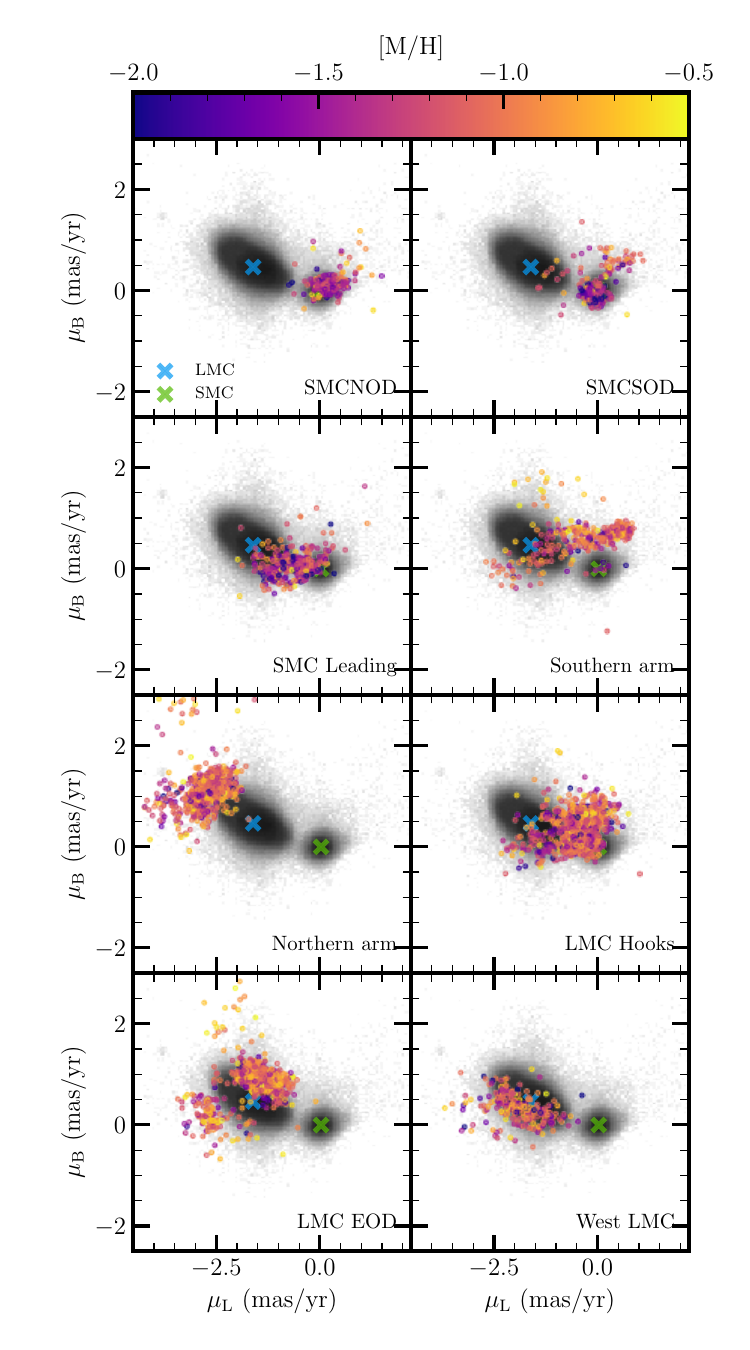}
    \caption{Proper motions in Magellanic Stream coordinates \protect\citep{Nidever2008} of the stars in our sample with respect to the SMC centre of mass. The stars of the main periphery structures of the Magellanic Clouds are represented in each of the panels, as indicated in Figure \ref{fig:periph_mdfs}. The grey scale background density corresponds to all the metallicity dataset out to 50 degrees from the LMC, and the blue and red crosses are the mean motions of each main body. The proper motions of the periphery features are colour coded according to their metallicity values.}
    \label{fig:periph_pms}
\end{figure}

\begin{figure}
    \centering
    \includegraphics[width=\columnwidth]{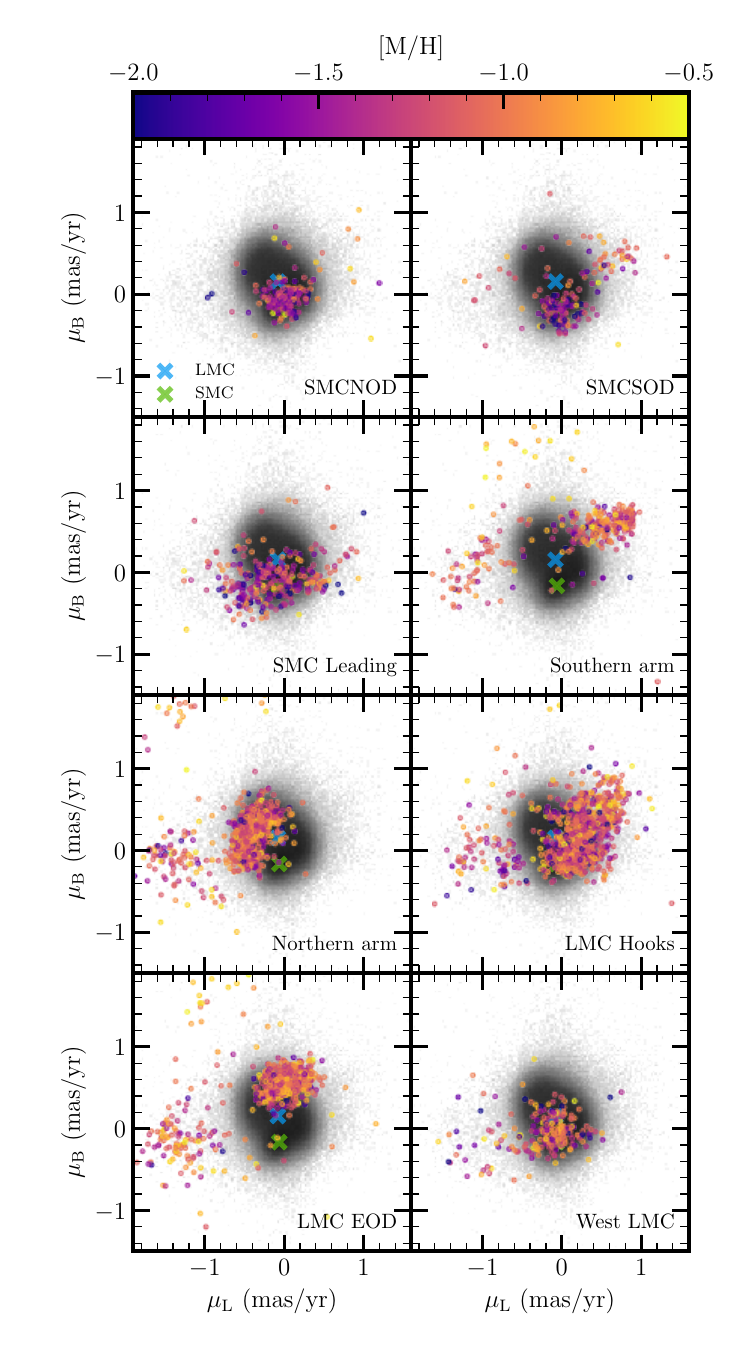}
    \caption{Same as Figure \ref{fig:periph_pms} but using the LMC as the center of mass reference.}
    \label{fig:periph_pms_lmc}
\end{figure}

In Figures \ref{fig:periph_pms} and \ref{fig:periph_pms_lmc} we show their motions in different panels and with the stars in each area colour coded by their metallicity. We depict in grey scale the motion for all stars in the sample with metallicities that were cross-matched with the \textit{Gaia} catalogue out to 50\dgr from the LMC centre, and with a metallicity up to -0.5 dex to enhance the kinematic signature of the outer features. For our discussion, we are going to focus on the SMC COM because it shows a more clear distinction between the different features as well as the MCs bulk motions.

In Figure \ref{fig:periph_pms}, the main density peak corresponds to the LMC, centred around $(\mu_{\mathrm{L}},\mu_{\mathrm{B}})=(-1.5\degr,0.5\degr)$, while the smaller density peak around $(\mu_{\mathrm{L}},\mu_{\mathrm{B}})=(0\degr,0\degr)$ corresponds to the SMC. The motions of the stars in the different structures are mostly in agreement with the picture depicted by the MDFs in Figure \ref{fig:periph_mdfs}. Of particular interest are the SMCSOD and Southern arm regions, which contain a portion of highly perturbed population. The SMCSOD region shows a clear double population, one with SMC-like motions and the other one clearly off-set with an apparent higher metallicity. We note that the distinction of two populations corresponds to the ``bulk'' and ``off-set'' populations in C23, which also showed that they have different line-of-sight (LOS) velocities. In this study, we add that the off-set population is in fact just a continuation from the Southern arm stars, based on their proper motions and metallicities. Additionally, the LMC Hooks also show a tidally perturbed component that overlaps in proper motion with the SMCSOD and Southern arm, and with a motion generally closer to the SMC COM. Figure \ref{fig:periph_pms_lmc}, with the LMC COM motion subtracted, is in agreement with these conclusions.

\section{Discussion}
\label{section:discussion}

Here we focus on what the different results in the previous section mean in the context of the formation and evolution of the Clouds. 

\subsection{Metallicity Gradients}
\label{subsec:discussion_gradients}

The metallicity gradients of the galaxy could be giving important clues about the history of the MCs.  Both MCs have striking radial metallicities gradients.  The two LMC metallicity plateaus are very evident in the metallicity map (Fig.\ \ref{fig:mhmap}). It is well-known that the outer LMC stellar populations are quite old and show very little age or metallicity spread \citep{Mackey2016,Nidever2019,Cheng2022,Munoz2023}.  In contrast, the inner LMC has a broad metallicity  \citep{Nidever2020a} and age \citep{Harris2009,Ruiz-Lara2020b} distribution. \citet{Monelli2011} determined star formation histories at various radii in the LMC and found that the young ($<$2 Gyr) and intermediate age (2--5 Gyr) populations have a steeper radial gradient than the older population ($>$8 Gyr) and that the youngest population drops off abruptly beyond 8\degr.  

These two spatial, metallicity and age regimes likely mark important phases in the evolution of the LMC.  The first is an early phase of evolution where the galaxy was quite extended but with a moderate metallicity ([Fe/H]$\sim -1$).  This was followed by a second phase of star formation that still continues today but has been more limited in its spatial extent.
In fact, there is evidence that the region of active star formation in the LMC (and, therefore, the gaseous component) has been continually decreasing in the last several Gyrs \citep{Gallart2008,Meschin2014} at a rate of $\sim$2.4 kpc Gyr$^{-1}$ and will be completely gone in just $\sim$0.8 Gyr \citep{Nidever2014}.  This ``whittling-away'' of the LMC's gaseous disk is likely due to ram pressure stripping from the MW and repeated and increasingly close interactions with the SMC. 

We propose that we call these two regions of the LMC the ``inner disk'' and the ``outer disk'', demarcated by the rapid drop in metallicity at around $\sim$8--9\dgr (although this radius depends on the position angle).  What could have caused such a dramatic change in the LMC's evolution 2--8 Gyr ago?
One hypothesis is that this is when the LMC and SMC first became bound (or loosely bound) together. There is evidence for the timing of this interaction based on star formation histories \citep{Massana2022}, and their first encounter likely would have disturbed both galaxies noticeably.
A second hypothesis is that the LMC made one previous passage of the MW and that this stripped off a decent amount of gas but left enough for the LMC to continue forming stars until the current time. Finally, it is possible that the outer disk stars were not born at these large radii, but rather, moved there by some tidal interaction in the distant past.
What ever the cause, the patterns identified here are clear observational signatures that can be used to constrain Magellanic models.
We note that there appear to be no star clusters in the outer disk of the LMC \citep{Bica2008}, that might provide some clues as to the origin of this feature. 

The SMC has a drop in its metallicity beyond $\sim$5\dgr which is where the clear elliptical shape of the SMC, as seen in the Gaia stellar distribution, ends.  The features beyond that are the SMCNOD, SMCSOD and the Leading populations.  These are all relatively diffuse and show a broad, and potentially bimodal, metallicity distribution (Fig.\ \ref{fig:periph_mdfs}).  These features are likely tidally stripped components from the SMC due to interactions with the LMC \citep{Besla2012,Diaz2012}. One line of evidence of tidal stripping are the two line-of-sight components in the eastern SMC \citep{Hatzidimitriou1989,Nidever2013} where the closer one appears to have been stripped from the inner SMC by the most recent interaction with the LMC $\sim$200 Myr ago.  More simulations of the MC interactions should help illuminate the origin of the outermost SMC features and their broad MDFs.

\subsection{The Origin of the Periphery} \label{subsec:periphery_discussion}

We find that combining the results obtained for the different periphery structures shown in the previous section, we can make several coherent hypotheses regarding their origin.

One of the more evident features of most of the debris identified in this work is its relatively higher metallicity. The debris around the SMC, namely the SMCNOD, SMCSOD and SMC Leading, are noticeably more metal poor ($\sim$0.2 dex lower than the other regions). Assuming that the stars in this area have been stripped more recently from the SMC, due to their proximity to the galaxy, it would seem unlikely that the higher metallicity debris also comes from the SMC itself. Only based on the [M/H] values, it would not be possible for stars to be further away from the SMC and at the same time be more metal rich. Therefore, we hypothesise that the rest of the stars are mostly stripped from the LMC due to the combined influence of the SMC and MW. The only exception being areas like the LMC Hooks, that have on average slightly lower metallicity ($\sim$0.1 dex) and therefore potentially have a small SMC contribution. 

This is an apparent contradiction to the popular belief that most of the debris should belong to the SMC \citep{Besla2013},
in the same way that the Magellanic Stream is thought to be SMC gas stripped off \citep{Besla2012,Lucchini2020}. However, later simulations have shown that the SMC is responsible for the perturbed arc-like structures in the LMC outskirts \citep{Besla2016}, which would explain more of the debris being stripped material from the LMC. The ``off-set'' population, from our SMCSOD region, discovered in C23 is attributed to the SMC in their most likely scenario. In our scenario, we show how these stars have a metallicity compatible with any other known LMC debris, such as the Northern and Southern arms (see Figure \ref{fig:periph_pms}, and in fact they most likely belong to the same structure as the Southern arm. The authors in C23 attributed the ``off-set'' population to stripped SMC stars from the central parts of the galaxy, due to the most recent interaction with the LMC \citep{Zivick2018}. Looking at the SMCSOD area in Figure \ref{fig:4density_maps}, the bottom two panels (more metal rich) do not appear to have a higher metallicity population, more than the metal poor top panels do. With this in mind, we find that stars in the LMC Hooks region are the only ones compatible with being recently tidally stripped due to the most recent LMC-SMC collision. With those areas having a well-connected feature to the SMC in the most metal rich regime. All other debris has likely been stripped in past interactions between them \citep{Massana2022}, or with the MW, and we are only observing what was stripped from the more massive galaxy (LMC) due to the debris being more dense to begin with. We speculate that SMC debris stripped a long time ago is already too spread out and not visible as a coherent structure. Although outside the scope of this paper, it would be beneficial to model the disruption of the SMC and calculate a lower limit on the time of a previous interaction that would make it so that the stars stripped then would not be visible today.

Furthermore, with the metallicity results provided in this study, we are able to confirm for the first time that the SMCNOD has a double population of stars using spectroscopic results. It is unclear how high metallicity stars have arrived at this area, given that their motion does not seem to be perturbed significantly from the SMC internal motions (i.e. both stellar groups follow the same motion as the SMC). Overall, this could indicate that the galaxy, as a whole, has been tightly coupled to the LMC for a significant period of its evolution while being loosely bound, and thus capable of shedding stars. Its tidal radius sitting around $\sim$4 kpc \citep{Besla2010,Massana2020}, would help affirm this hypothesis.

\section{Conclusions}
\label{section:conclusions}

We have used metallicities derived from the \textit{Gaia} low-resolution XPSpec data to study the stellar structures of the Magellanic Clouds. We perform a detailed decontamination of MW foreground stars using the extra information given by surface gravity values for each star. This yields a detailed metallicity map of the Magellanic Clouds, that we use to study its metallicity gradients and periphery structures.

We find that the metallicity gradient of the LMC and SMC plateaus at intermediate radii, between 3-7$^{\circ}$ for the LMC and 3-5$^{\circ}$ for the SMC. At larger radii, the gradient for the LMC stabilises at a lower metallicity value, while the SMC gradient increases. We speculate that this is due to mixed LMC debris around the galaxy.

We identify eight different areas with known Magellanic debris in the outskirts of the galaxies to study their origins through their metallicities. The MDFs for these regions reveal most of them to be compatible with being dominated by LMC debris, while only the areas immediately next to the SMC show a clear distinction in metallicity, being more metal poor. The distribution of debris in different metallicity bins show that most of the population is concentrated around the $-1.3<\mathrm{[M/H]}<-0.9$ range, compatible with LMC debris. The SMC shows a more prominent debris structure at a more metal poor regime, but still has some metal rich debris around that could be stripped from the LMC.

Additionally, we perform a kinematic analysis of the different structures with respect to the COM motion of the SMC. Interestingly, we show that the motion of the Southern Arm and points to the same direction of its MDF and shows not to be directly connected with the SMC, although we note it seems to be a very perturbed region compared to others. We also identify a small component of the Southern arm 
overlapping SMC debris in the SMCSOD region.
This had originally been attributed to the SMC, but we favour a relation directly to the LMC based on metallicity.

Using the capabilities of new \textit{Gaia} spectroscopic data, we were capable of making an accurate characterisation of all the periphery structures that have been detected so far using much deeper photometry. This shows the capability of physical parameters to further aid in membership selection of MC stars. Additionally, metallicity estimates for more structures in the Clouds can further help constrain origins for them and the interactions that caused them. As we have seen here, even with the current depth we can already challenge our current understanding of their formation.  The addition of radial velocities from higher-resolution spectroscopic results (e.g, MagES, SDSS/APOGEE and 4MOST) will provide even more information on the origin and orbits of the stellar debris.


\section*{Acknowledgements}

The authors thank the referee for helpful comments that helped improve the manuscript.

P.M. acknowledges support from the National Science Foundation (AST-1908331).
D.L.N. acknowledges support from the National Science Foundation (AST-2109196).

Computational efforts were performed on the Tempest High Performance Computing System, operated and supported by University Information Technology Research Cyberinfrastructure at Montana State University.

This work has made use of data from the European Space Agency (ESA) mission {\it Gaia} (\url{https://www.cosmos.esa.int/gaia}), processed by the {\it Gaia} Data Processing and Analysis Consortium (DPAC, \url{https://www.cosmos.esa.int/web/gaia/dpac/consortium}). Funding for the DPAC has been provided by national institutions, in particular the institutions participating in the {\it Gaia} Multilateral Agreement.

\section*{Data Availability}

The \textit{Gaia} data used in this paper comes from the third data release and can be accessed from \url{https://gea.esac.esa.int/archive/}.

The physical parameters used in this paper have been obtained through \url{https://doi.org/10.5281/zenodo.7599788} and are publicly available.

The results presented in this paper are available upon reasonable request to the corresponding author.



\bibliographystyle{mnras}
\bibliography{references}




\appendix

\section{Individual MDFs for periphery regions}

\begin{figure}
    \centering
    \includegraphics[width=\columnwidth]{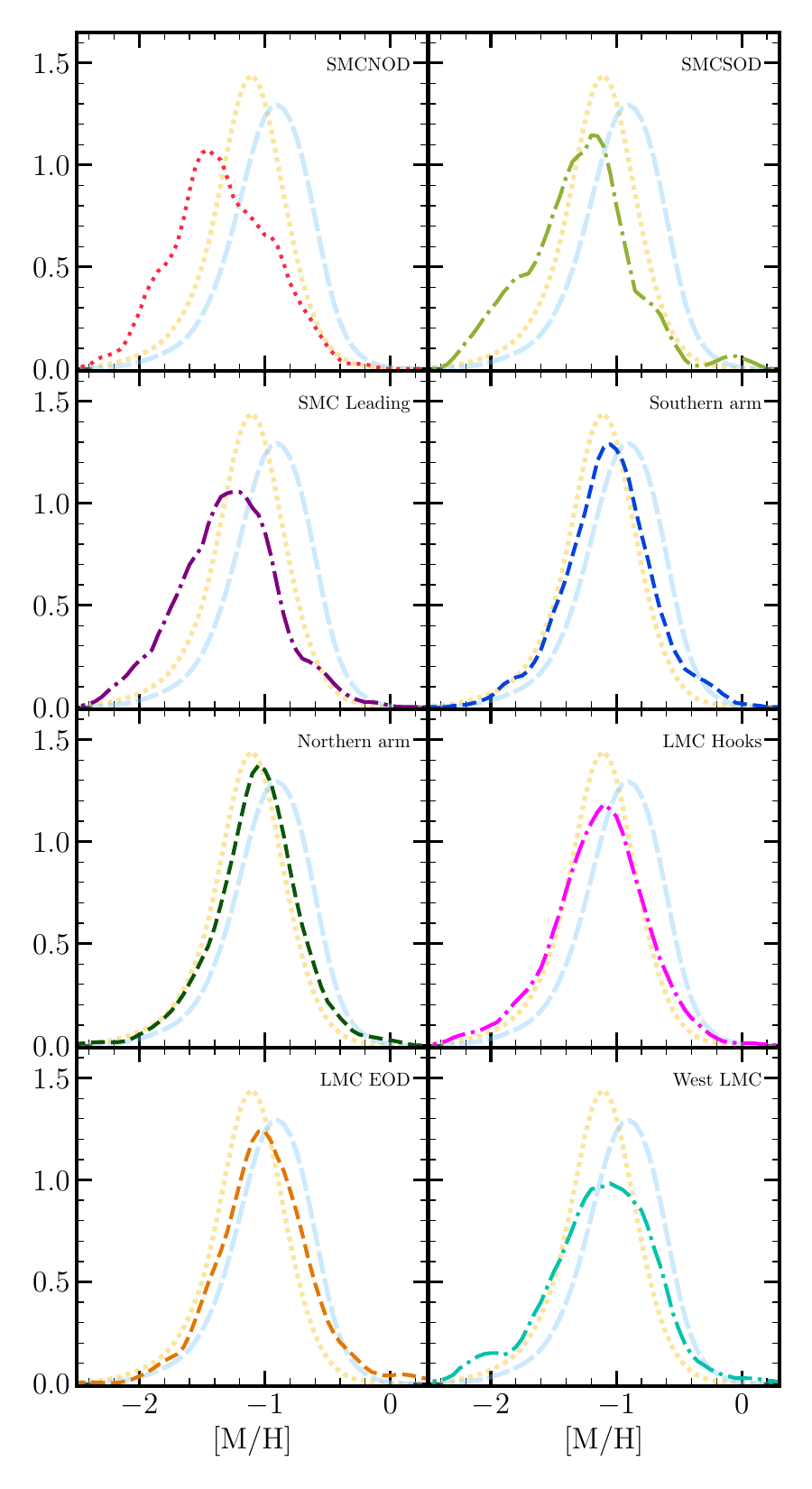}
    \caption{Metallicity distribution functions for each periphery region highlighted in Figure \ref{fig:periph_mdfs}. Each panels also shows the MDFs of SMC (yellow dotted) and LMC (blue dashed), for comparison.}
    \label{fig:single_mdfs}
\end{figure}

In this Appendix we want to show the MDFs presented in the bottom panels of Figure \ref{fig:periph_mdfs} by themselves, without having them overlap with each other. In order to keep a visual reference, we keep the LMC and SMC main body MDFs in all panels. This is shown in Figure \ref{fig:single_mdfs}.

\bsp	
\label{lastpage}
\end{document}